\documentclass[12pt]{iopart}
\usepackage{graphicx}
\usepackage{latexsym}
\usepackage{amssymb}
\usepackage{bm}

\newcommand{\mathsym}[1]{{}}

\newcommand{\f}{\begin{equation}}
\newcommand{\ff}{\end{equation}}
  \def\msol{\ifmmode M_\odot\else$M_\odot$\fi}
  
\begin{document}

%\title{Characteristics of Relativistic MHD Turbulence}
\title{Invariants in Relativistic MHD Turbulence}

\author{David Garrison}
\address{Physics Department, University of Houston Clear Lake, Houston, Texas 77058}
\ead{garrison@uhcl.edu}

\author{Phu Nguyen}
\address{Physics Department, University of Houston Clear Lake, Houston, Texas 77058}
\ead{nguyenp@uhcl.edu}

%\date{\today}

\begin{abstract}
The objective of this work is to understand how the characteristics of relativistic MHD turbulence may differ from those of nonrelativistic MHD turbulence.  We accomplish this by studying the ideal invariants in the relativistic case and comparing them to what we know of nonrelativistic turbulence.  Although much work has been done to understand the dynamics of nonrelativistic systems (mostly for ideal incompressible fluids), there is minimal literature explicitly describing the dynamics of relativistic MHD turbulence using numerical simulations.  Many researchers simply assume that relativistic turbulence has the same invariants and obeys the same dynamics as non-relativistic systems our results show that this assumption may be incorrect.
\end{abstract}

%\keywords{Cosmology: Numerical Relativity: Relativistic Plasma}

%\pacs{98.80.-k 04.25.D- 52.27.Ny}

\maketitle

\section{Introduction}

Many studies in numerical relativity and high-energy astrophysics depend on the dynamics of relativistic plasmas.  These include phenomenon such as primordial turbulence, neutron stars, active galactic nuclei, and accretion disks near black holes~\cite{Caprini:2006jb,Caprini:2009yp,cho2006,garrison,inoue,Melatos:2009mz,zhang,zrake}.  Unfortunately, we do not know if the results of these studies are accurate because of approximations such as the use of nonrelativistic fluid dynamics and the lack of a standard model to describe the dynamics of relativistic turbulence.  In particular, very little is understood about the turbulent dynamics of a relativistic plasma or its effect on the evolution of magnetic fields. This can only effectively be studied through direct numerical simulation of the relativistic magnetofluid.

In the following report we will first discuss what is currently known about the dynamics of nonrelativistic MHD systems.  We introduce the standard incompressible and compressible nonrelativistic MHD evolution equations as well as the ideal invariants for those systems.  In section 4, we will introduce the relativistic MHD equations and the relativistic equivalents of the nonrelativistic ideal MHD invariants.  We then describe our numerical experiment and present our results for a relativistic MHD code.  We conclude by discussing the similarities and differences between the different systems.

\section{Nonrelativistic Incompressible MHD}

Work by Shebalin \cite{shebalin1988,shebalin}, on ideal homogeneous incompressible MHD turbulence best demonstrates how the dynamics of a magnetofluid can differ from that of a hydrodynamic fluid.  Plasma can be accurately modeled as a fluid made up of charged particles that are therefore affected by magnetic fields as well as particle-particle interactions.  Because of this, the magnetic field becomes a dynamic variable in addition to density, pressure and the velocity of particles.  For example, In MHD turbulence, an equipartition occurs and we expect kinetic and magnetic energy fluctuations to become roughly equal.  Shebalin modeled the magnetofluid as a homogenous system where the same statistics are considered valid everywhere in the computational domain.  He utilized periodic boundary conditions and spectral methods in order to study how the dynamics of different scales interacted without the addition of boundary errors.  Much of his work focused on an ideal MHD system, where the magnetic and fluid dissipation terms were excluded.  Below are the evolution equations used by Shebalin to describe the incompressible MHD system.
\numparts
\begin{eqnarray}
\vec{\nabla} \cdot \vec{v} &= 0      \\
\rho \frac{\partial \vec{v}}{\partial t} &= -\rho (\vec{v} \cdot \nabla v) - \vec{\nabla} p - \frac{1}{4 \pi} \vec{B} \times ( \vec{\nabla} \times \vec{B} ) \\
\frac{\partial \vec{B}}{\partial t} &= \vec{\nabla} \times (\vec{v} \times \vec{B}).
\end{eqnarray}
\endnumparts
By varying the mean magnetic field ($B_{0}$) and angular velocity ($\Omega_{0}$) of the system, Shebalin was able to define five different cases with different invariants as shown in Table \ref{tab:MHDtests1}. In such a system there could be as many as 3 ideal invariants; energy ($E$), and the psuedoscalars cross helicity ($H_{C}$) and magnetic helicity ($H_{M}$).  In addition, the invariant parallel helicity, $H_{P} = H_{C} - \sigma H_{M}$  $(\sigma = \Omega_{0}/B_{0})$, can be formed from a linear combination of cross and magnetic helicity. In a hydrodynamic fluid the ideal invariants are only energy and kinetic helicity.

\begin{table*}
\caption{Invariants for Ideal Incompressible MHD. \label{table:MHD1}}
\centering
\begin{tabular}{c  c  c  c}
\hline \hline
 Case & Mean Field & Angular Velocity & Invariants \\ 
\hline \hline
I &	0 &	0 &	$E, H_{C}, H_{M}$ \\
\hline
II &	$B_{0} \ne 0$ &	0 &	$E, H_{C}$ \\
\hline
III &	0 &	$\Omega_{0} \ne 0$ &	$E, H_{M}$ \\
\hline
IV &	$B_{0} \ne 0$ &	$\Omega_{0} = \sigma B_{0}$ &	$E, H_{P}$  \\
\hline
V &	$B_{0} \ne 0$ &	$\Omega_{0} \ne 0$ ($B_{0} \times \Omega_{0} \ne 0$ ) & $E$  \\
\hline \hline
\end{tabular}
\vskip 12pt
\label{tab:MHDtests1}
\end{table*}

For an incompressible fluid u(k,t) is the Fourier coefficient of turbulent velocity and b(k,t) is the Fourier coefficient of the turbulent magnetic field.  The energy, cross helicity and magnetic helicity can be expressed in terms of these as:
\numparts
\begin{eqnarray}
E & =  \frac{1}{2 N^3} \sum_k [ |u(k)|^2 + |b(k)|^2]      \\
H_{C} & = \frac{1}{2 N^3} \sum_k u(k) \cdot b^{*}(k)  \\
H_{M} & = \frac{1}{2 N^3} \sum_k \frac{i}{k^2} k \cdot b(k) \times  b^{*}(k).
\end{eqnarray}
\endnumparts
A cubic computational domain with N grid points in each direction is assumed.  The statistical mechanics of the system is defined by the Gaussian canonical probability density function (PDF):
\numparts
\begin{eqnarray}
D & =  \frac{1}{Z} exp(-\alpha E - \beta H_{C} - \gamma H_{M} ) \\
Z & =  \int_{\Gamma} exp(-\alpha E - \beta H_{C} - \gamma H_{M} ) d \Gamma \\
\hat{\Phi} & = \int_{\Gamma} \Phi D d \Gamma, ~~~ \bar{\Phi} = \frac{1}{T} \int_{0}^{T} \Phi dt.
\end{eqnarray}
\endnumparts
Where Z is the partition function and d$\Gamma$ is the phase space volume.  $\hat{\Phi}$ shows how to calculate the ensemble averages using the PDF while $ \bar{\Phi}$ is the time average.  If $\hat{\Phi} = \bar{\Phi}$, the system is said to be ergodic but if $\hat{\Phi} \ne \bar{\Phi}$, it is non-ergodic.  Here $\alpha$, $\beta$ and $\gamma$ are inverse temperatures.  The ensemble average magnetic energy ($\hat{E}_M$) is always greater than or equal to ensemble average kinetic energy ($\hat{E}_K$), and the inverse temperature terms can be found as a function of $\hat{E}_M$ = $\phi$.
\numparts
\begin{eqnarray}
\alpha & =  \frac{2 N^3 \phi}{\phi (\hat{E} - \phi) - \hat{H}_{C}^2} \\
\beta & =  -2 \frac{\hat{H}_{C}}{\phi} \alpha \\
\gamma & =  - \frac{2 \phi - \hat{E}}{\hat{H}_{M}} \alpha
\end{eqnarray}
\endnumparts
Phase portraits resulting from computer simulations of Shebalin$\textquoteright$s five cases show that coherent structures formed in many systems where the magnetofluid was experiencing turbulence \cite{shebalin1988,shebalin}.  Coherent structure occurs when time-averaged physical variables in MHD turbulence have large mean values, rather than the zero mean values expected from theoretical ensemble predictions. MHD turbulence thus has  Òbroken ergodicityÓ, which can be explained by finding the eigenmodes of the system. One out of the four eigenvalues associated with each of the lowest wavenumbers will be very much smaller than the others; the eigenvariables associated with these very small eigenvalues grow to have very large energies compared to other eigenvariables; when this happens, an almost Òforce-free stateÓ occurs in which large-energy eigenmodes are quasistationary while low-energy eigenmodes remain turbulent; thus, the predicted ergodicity has been dynamically broken. This is observed to occur even in dissipative systems because broken ergodicity in MHD turbulence manifests itself at the smallest wavenumbers (largest length scales) where dissipation is negligible, resulting in the ideal spectrum. In the case of ideal hydrodynamic turbulence, broken ergodicity can occur in a finite model system, but only at the largest wavenumbers (smallest scales).  When dissipation is added, the large wavenumber modes are most affected and their energy quickly decays away, so that broken ergodicity plays no role in decaying hydrodynamic turbulence.  

\section{Nonrelativistic Compressible MHD}

Compressible MHD systems have not been studied as much as incompressible systems so here we will focus primarily on their invariants.  We will assume that both incompressible and compressible systems share the same statistical mechanics and dynamics whenever the same invariants apply.  In a nonrelativistic compressible MHD system; Energy and the Incompressible form of Cross Helicity are always conserved for a nondissipative system~\cite{Banerjee, Marsch,yokoi} (See Table 2).  Compressible Cross Helicity, $\tilde{H}_{C} = \rho H_{C}$, is not an ideal invariant in compressible MHD~\cite{Banerjee}. In the absence of a mean magnetic field and dissipation, Magnetic Helicity is also conserved~\cite{Ghosh}.  The authors were unable to identify any literature showing the relationship between net angular velocity and ideal compressible MHD invariants. 

Given that our relativistic system is by default a compressible system, we naively expect to see that the same ideal invariants will apply for the relativistic system as the nonrelativistic compressible system. The equations for ideal compressible MHD are similar to those of the incompressible system with the exception of the first equation.
\numparts
\begin{eqnarray}
\frac{\partial \rho}{\partial t} &=  -\vec{\nabla} \cdot (\rho \vec{v})  \\
\rho \frac{\partial \vec{v}}{\partial t} &= -\rho (\vec{v} \cdot \nabla v) - \vec{\nabla} p - \frac{1}{4 \pi} \vec{B} \times ( \vec{\nabla} \times \vec{B} )  \\
\frac{\partial \vec{B}}{\partial t} &= \vec{\nabla} \times (\vec{v} \times \vec{B}).
\end{eqnarray}
\endnumparts
\begin{table*}
\caption{Invariants for Ideal Compressible MHD. \label{table:MHD2}}
\centering
\begin{tabular}{c  c  c  c}
\hline \hline
 Case & Mean Field & Invariants \\ 
\hline \hline
I &	0 & $E, H_{C}, H_{M}$ \\
\hline
II &	$B_{0} \ne 0$ & $E, H_{C}$ \\
\hline \hline
\end{tabular}
\vskip 12pt
\label{tab:MHDtests}
\end{table*}

\section{Relativistic MHD Systems}

The fluid and electromagnetic components of the relativistic MHD equations are developed from several well-known equations~\cite{duez}. They include the conservation of particle number, the continuity equation, the conservation of energy-momentum, the magnetic constraint equation and the magnetic induction equation. For a system consisting of a perfect fluid and an electromagnetic field, the ideal MHD stress-energy tensor is given by \numparts
\begin{eqnarray}
T^{\mu\nu}  & = (\rho_0 h + b^{2})u^{\mu}u^{\nu} +\bigl(P + \frac{b^{2}}{2}\bigr)g^{\mu\nu} - b^{\mu}b^{\nu}\label{eq:s-e} \\
P &= \epsilon \rho_0 (\tilde{\Gamma} -1) \\
h & = 1 + \epsilon + \frac{P}{\rho_0}\, \\
b^{\mu} & = \frac{1}{\sqrt{4\pi}}B^{\mu}_{(u)}\, \\
B^{0}_{(u)} & = \frac{1}{\alpha}u_i B^{i}\, \\
B^{i}_{(u)} & = \frac{1}{u^{0}}(\frac{B^{i}}{\alpha} + B^{0}_{(u)} u^i)\,.
\end{eqnarray}
\endnumparts
Here, $P$ is the fluid pressure, $\rho_0$ is density, $B^i$ is magnetic field, $u^\mu$ is four-velocity, $h$ is the enthalpy, $\epsilon$ is specific internal energy, and $b^2$ is the magnitude of the magnetic vector field squared.  We define pressure in terms of the energy density using the $\tilde{\Gamma}$ law equation of state with, $\tilde{\Gamma} = \frac{5}{3}$.  $P = (\tilde{\Gamma} - 1) \rho \epsilon$ at most energies and $P \propto \rho^{\tilde{\Gamma}}$ at extremely low energies.  The evolution equations where given by Duez as~\cite{duez}:
\numparts
\begin{eqnarray}
\partial_t \rho_* &= - \partial_j (\rho_* v^j) \\
\partial_t \tilde{\tau} &= - \partial_i (\alpha^2 \sqrt{\gamma} T^{0 i} - \rho_* v^i) + s \\
\partial_t \tilde{S_i} &= - \partial_j (\alpha \sqrt{\gamma} T^j_i) - \frac{1}{2} \alpha \sqrt{\gamma} T^{\alpha \beta} g_{\alpha \beta, i} \\
\partial_t \tilde{B}^i &= - \partial_j (v^j \tilde{B}^i - v^i \tilde{B}^j).
\end{eqnarray}
\endnumparts
Here, $\rho_{*}$ is conserved mass density, $\tilde{\tau}$ relates to energy density,  $ \tilde{S_i}$ is momentum density, $\tilde{B}^i$ is related to the magnetic field and s is the source term.  $\gamma_{ij}$ is the three metric and $\alpha$ is a lapse term related to the time evolution of the simulation.  The determinate of the three metric and lapse are both set to unity because we are using the Minkowski metric and Geodesic slicing conditions.
\numparts
\begin{eqnarray}
\rho_* &= \alpha \sqrt{\gamma} \rho_0 u^0 \\
\tilde{\tau} &= \alpha^2 \sqrt{\gamma} T^{00} - \rho_* \\
\tilde{S_i} &= \alpha \sqrt{\gamma} T^{0}_{i} = (\rho_* h + \alpha u^0 \sqrt{\gamma} b^2) u_i -  \alpha \sqrt{\gamma} b^0 b_i \\
\tilde{B}^j &= \sqrt{\gamma} B^j \
\end{eqnarray}
\endnumparts
Notice that unlike the nonrelativistic system, we use the stress-energy tensor within the evolution equations so that 4-momentum conservation is built into the system.  This results in a set of equations that look very different from that of the nonrelativistic system.  According to work by Yoshida \cite{Yoshida, Kawazura}, in addition to 4-momentum, relativistic systems are expected to conserve a quantity called Relativistic Helicity.  It is defined below using the canonical 4-momentum density, $\mathcal{P}$, of the system.
\begin{eqnarray}
\bf{\kappa} = (\bf{\mathcal{P}} \cdot (\nabla \times \bf{\mathcal{P}}), \mathcal{P}_{0} (\nabla \times \bf{\mathcal{P}}) + \bf{\mathcal{P}} \times (\nabla \mathcal{P}_{0} + \partial_0 \bf{\mathcal{P}})) 
\end{eqnarray}
Here the canonical 4-momentum density $\mathcal{P} = (\mathcal{P}_{0},\bf{\mathcal{P}})$ is a combination of mechanical and electromagnetic momentum densities.  The conservation of Relativistic Helicity is then effectively, $\int \partial_{\mu} \bf{\kappa^{\mu}} d^3x= 0$.  The canonical 4-momentum can be expressed as the sum of mechanical and electromagnetic momentum, $\mathcal{P}_{\mu} = P_{\mu}  + e \mathcal{A}_{\mu} $.  If we ignore the electromagnetic fields, we recover a relativistic version of Cross Helicity Density, $\bf{\kappa_C}$.  If we set the particle's mechanical momentum to zero, we recover a relativistic version of Magnetic Helicity Density, $\bf{\kappa_{M}}$.  

\section{Methodology}

For this numerical experiment, we calculate Energy Density (E), Relativistic Helicity Density ($\bf{\kappa}$), Cross Helicity Density ($\bf{\kappa_C}$), and Magnetic Helicity Density ($\bf{\kappa_{M}}$), numerically using the code described later in this section.  These variables are defined as shown below.
\numparts
\begin{eqnarray}
E  & =  \int T^{00}(\vec{x}) d^3x\\
\bf{P}_0  &= \rho_{*} h u_0 - P \\
\bf{P}_{i} &= \rho_{*} h u_i \\
\bf{\kappa} &= ( \bf{\tilde{S}} \cdot (\nabla \times  \bf{\tilde{S}}), (\tau + \rho_{*}) (\nabla \times \bf{\tilde{S}}) + \bf{\tilde{S}} \times (\nabla (\tau + \rho_{*}) + \partial_0 \bf{\tilde{S}}))  \\
\bf{\kappa_C} &=  (\bf{P} \cdot B, P_0 B - \bf{P} \times \bf{\mathcal{E}}) \\
\bf{\kappa_{M}} &= (\bf{\mathcal{A}}  \cdot \bf{\mathcal{B}}, \mathcal{A}_{0} \bf{\mathcal{B}} - \bf{\mathcal{A}} \times \bf{\mathcal{E}}) 
\end{eqnarray}
\endnumparts
Here the magnetic field is related to the vector potential by the equation, $\bf{\mathcal{B}} = \nabla \times \bf{\mathcal{A}}$.  The electric field is defined using the MHD conditions, $\bf{\mathcal{E}} = \bf{\mathcal{B}} \times \bf{v}$.  We can test to see if the Helicities are invariant by comparing numerically calculated time derivatives of $\kappa^0$ to their predicted value at each time-step using the equation $\partial_0 \bf{\kappa^0} + \partial_i \bf{\kappa^i} = 0$.  We normalize the result using the L2 norm of the calculated divergence so all results are on the same relative scale.  We then integrate the result over the volume of our computational domain.  If the normalized error is dominated by the truncation and round-off errors, we can assume that the system is invariant.  For the normalized error in energy we simply look at the difference in energy at two different time levels divided by the total energy at that time level.
\numparts
\begin{eqnarray}
Error_{\kappa}  & = \frac{\int [(\bf{\kappa^0}(t) - \bf{\kappa^0}(t - \Delta t))/ \Delta t + \partial_i \bf{\kappa^i}(t)] d^3x}{\int ||\partial_i \bf{\kappa^i}|| d^3x} \\
Error _{E} &= \frac{E(t) - E(t - \Delta t)} {E(t)}
\end{eqnarray}
\endnumparts
In order to study the invariants of the relativistic MHD equations, we used a code called FixedCosmo which was originally written by one of us~\cite{garrison} to study the dynamics of primordial plasma turbulence.  This code was developed using the open source Cactus framework (www.cactuscode.org).  Cactus was originally developed to perform numerical relativistic simulations of colliding black holes but it's modular design has since allowed it to be used for a variety of Physics, Engineering and Computer Science applications.  It is currently being maintained by the Center for Computation and Technology at Louisiana Sate University.  Cactus codes are composed of a flesh (which provides the framework) and the thorns (which provide the physics).  FixedCosmo is a collection of thorns.  It uses the form of the Relativistic MHD equations described by Duez~\cite{duez} and is written in a combination of F77, F90, C and C++.  This code is parallelized and capable of using several different differencing methods such as second order finite differencing, fourth order finite differencing and spectral differencing.  Although the code is capable of utilizing artificial viscosity and HRSC, neither was used for this project.

Because the objective of this study is to test the ideal relativistic MHD system, we complete a series of runs in a ``high-energy'' regime.   The parameters used approximate that of the early universe around the electroweak scale.  This is done so we can apply the results to any relativistic MHD system.  Table 3 shows a matrix of the test runs.

\begin{table}[ht]
\caption{High Energy Numerical Simulations}
\centering                                       
\begin{tabular}{c c c c c c}               
\hline
Variables & Case 1 & Case 2 & Case 3 & Case 4 & Case 5 \\[0.2ex] 
%heading
\hline                                    
Max Velocity (c) & 0.25 & 0.25 & 0.25 & 0.25 & 0.25\\                       
$\Omega_x$ (rad/s) & 0 & 0 & 0 & 0 & 0\\
$\Omega_y$ (rad/s) & 0 & 0 & 0 & 0 & 0\\
$\Omega_z$ (rad/s) & 0 & 0 & 0.35 & 0.35 & 0.35\\
Init Temperature (K) & 2.8e15 & 2.8e15 & 2.8e15 & 2.8e15 & 2.8e15\\
Init Density ($kg/m^{3}$) & 9.7e29 & 9.7e29 & 9.7e29 & 9.7e29 & 9.7e29\\
Max Magnetic Field (G) & 1.0e13 & 1.0e13 & 1.0e13 & 1.0e13 & 1.0e13\\
$B_x$ (G) & 0 & 2.0e13 & 0 & 0 & 2.0e13\\
$B_y$ (G) & 0 & 0 & 0 & 0 & 0\\
$B_z$ (G) & 0 & 0 & 0 & 2.0e13 & 0\\[1ex]
           
\hline                                
\end{tabular}
\label{table:HE}                
\end{table}

Each data run utilized Fourier spectral differencing on a grid with 64 x 64 x 64 internal data points.  We ran these simulations for about 7500 iterations or over $10^{-9}$ s of physical time.  The electron oscillation time for the ``high energy'' regime is about $1.1 \times 10^{-34}$ s so the simulations appear to have run long enough to witness the full dynamics of the system.  The simulation domains where all set to 4m x 4m x 4m.  Also, the code utilizes geometerized units ($c = G = \hbar = 1$) so parameters where translated from SI units to units of length for the use of the calculations and then back to SI units for the output.  Time is therefore translated into seconds by dividing the output time by the speed of light. 

\section{Results}

Truncation errors were found by doubling the resolution and measuring the change in the observed total errors.  By assuming that the Euler Method, used to calculate $\partial_0 \kappa^0$, is first order and that the numerical errors and variations from exactly conserved values are additive, the truncation errors can be calculated from 
\begin{eqnarray}
|E_{truncation}| = 2 (|E_{LR}| - |E_{HR}|).
\end{eqnarray}
If the truncation errors are within an order of magnitude of the normalized errors, we can conclude that the system is invariant.  We also found it impossible to completely eliminate the mean magnetic field and mean angular momentum in all cases.  A mean magnetic field (on the order of 1\% of the maximum field) remained in every case.  Also, each case seemed to have a small angular velocity, also less than 1\% of the fluid velocities within the simulation.  The authors feel that these residual quantities where not enough to significantly disrupt the system and could be safely ignored.

\begin{table}[ht]
\caption{High Energy Simulation Results}
\centering                                       
\begin{tabular}{c c c c c}               
\hline
Mean Magnitude of Errors & E & $\kappa$ & $\kappa_{C}$ & $\kappa_{M}$  \\[0.2ex] 
%heading
\hline                                    
Case 1 & 1.8e-9  &  2.8e-14 & 9.6e-5 & 1.0e-2 \\                      
Case 2 & 1.8e-9 & 2.8e-14 & 9.7e-5 & 2.8e-2 \\
Case 3 & 8.1e-10 & 0.0 & 6.6e-5 & 7.1e-3 \\
Case 4 & 7.7e-10 & 0.0 & 6.0e-5 & 1.9e-2 \\
Case 5 & 7.8e-10 & 0.0 & 6.6e-5 & 1.8e-2 \\
Truncation Errors & 2.1e-9 & 2.2e-14 & 2.9e-5 & 2.0e-3 \\
\hline                                
\end{tabular}
\label{table:HE}                
\end{table}

The results in Table 4 were calculated by averaging the absolute values of normalized errors. As one can see, Energy and Relativistic Helicity appear to be conserved in every case given that the normalized error appears to be dominated by truncation errors for both.  Cross Helicity may be conserved in every case since the calculated truncation error appears to be within an order of normalized errors in every case.  Magnetic Helicity does not appear to be conserved in any of the cases.  Normalized Errors for Cross Helicity Conservation appear smaller in cases where a large mean angular velocity is present.  Deviations in Magnetic Helicity Conservation are smallest in the absence of a large mean magnetic field. 

\section{Discussion}

Our results show that in the high-energy Relativistic MHD regime only Energy and Relativistic Helicity are clearly conserved.  We are not able to conclusively prove Cross Helicity conservation.  Magnetic Helicity conservation is questionable in this system.  This is not an unexpected result but it does raise several interesting questions which lie beyond the scope of this article.  Does the potential lack of Cross and Magnetic Helicity Conservation effect the dynamics of the relativistic system when it comes to phenomena such as inverse Energy Cascade or the Kolmogorov Energy Spectrum?  How do magnetic dynamos in relativistic MHD systems function?  Are there any other overlooked dynamics in relativistic MHD systems?  These are all questions which we hope to address in future numerical studies.

\section{Conflicts of Interests}

The authors declare that there is no conflict of interests regarding the publication of this article.

\section{Acknowledgement}

The authors would like to acknowledge the support of the University of Houston Center for Advanced Computing and Data Systems for access to the high performance computing resources used for the completion of this project.  The authors would also like to thank John Shebalin for several useful conversations and helpful suggestions.

\end{document}